\documentstyle[12pt]{article}
\begin{document}

\author{C. Bizdadea \thanks{%
e-mail address: bizdadea@central.ucv.ro} \\
Faculty of Physics, University of Craiova\\
13 A. I. Cuza Str., Craiova RO-1100, Romania}
\title{Note on two-dimensional nonlinear gauge theories}
\maketitle

\begin{abstract}
A two-dimensional nonlinear gauge theory that can be proposed for
generalization to higher dimensions is derived by means of cohomological
arguments.

\end{abstract}

A big step in the progress of the BRST formalism was its cohomological
understanding \cite{1}, which allowed, among others, a useful investigation
of many interesting aspects related to the perturbative renormalization
problem \cite{4}--\cite{5}, anomaly-tracking mechanism \cite{5}--\cite{6},
simultaneous study of local and rigid invariances of a given theory \cite{7}%
, as well as to the reformulation of the construction of consistent
interactions in gauge theories \cite{7a} in terms of the deformation theory 
\cite{8}, or, actually, in terms of the deformation of the solution to the
master equation. Joint to these topics, the problem of obtaining consistent
deformations has naturally found its extension at the Hamiltonian level by
means of local BRST cohomology \cite{11}. There is a large variety of models
of interest in theoretical physics that have been investigated in the light
of the deformation of the master equation \cite{9}--\cite{10}.

In this paper we investigate the consistent interactions that can be added
among a set of scalar fields, two types of one-forms and a system of
two-forms in two dimensions, described in the free limit by an abelian BF
theory \cite{12}, in order to construct a two-dimensional nonlinear gauge
theory that can be proposed for generalization to higher dimensions.
Nonlinear gauge theories \cite{13} are important as they include pure
two-dimensional gravitation theory \cite{14}, which is expected to offer a
conceptual mechanism for the study of quantum gravity in higher dimensions
from the perspective of gauge theories. More precisely, when the nonlinear
algebra is the Lorentz-covariant extension of the Poincar\'{e} algebra, the
theory turns out to be the Yang-Mills-like formulation of ${\bf R}^{2}$
gravity with dynamical torsion, or generic form of `dilaton' gravity \cite
{15}.

Our strategy goes as follows. Initially, we determine the antifield-BRST
symmetry of the free model, that splits as the sum between the Koszul-Tate
differential and the exterior derivative along the gauge orbits, $s=\delta
+\gamma $. Next, we deform the solution to the master equation of the free
model. The first-order deformation belongs to $H^{0}(s|d)$, where $d$ is the
exterior space-time derivative. The computation of the cohomological space $%
H^{0}(s|d)$ proceeds by expanding the co-cycles according to the antighost
number, and by further using the cohomological spaces $H(\gamma )$ and $%
H_{2}(\delta |d)$. Subsequently, we show that the consistency of the
first-order deformation requires that all the higher-order deformations
vanish. With the help of the deformed solution to the master equation we
finally identify the interacting theory and its gauge transformations, which
form a nonlinear gauge algebra (open algebra) that only closes on-shell.

We begin with a free model given by an abelian two-dimensional BF theory
involving a set of scalar fields, two types of one-forms and a system of
two-forms 
\begin{equation}
S_{0}\left[ A_{\mu }^{a},H_{\mu }^{a},\varphi _{a},B_{a}^{\mu \nu }\right]
=\int d^{2}x\left( H_{\mu }^{a}\partial ^{\mu }\varphi _{a}+\frac{1}{2}%
B_{a}^{\mu \nu }\partial _{\left[ \mu \right. }A_{\left. \nu \right]
}^{a}\right) ,  \label{1}
\end{equation}
subject to the irreducible gauge invariances 
\begin{equation}
\delta _{\epsilon }A_{\mu }^{a}=\partial _{\mu }\epsilon ^{a},\;\delta
_{\epsilon }H_{\mu }^{a}=\partial ^{\nu }\epsilon _{\mu \nu }^{a},\;\delta
_{\epsilon }\varphi _{a}=0,\;\delta _{\epsilon }B_{a}^{\mu \nu }=0,
\label{2}
\end{equation}
where the notation $[\mu \nu ]$ means antisymmetry with respect to the
indices between brackets. A consistent deformation of the free action (\ref
{1}) and of its gauge invariances (\ref{2}) defines a deformation of the
corresponding solution to the master equation that preserves both the master
equation and the field/antifield spectra. So, if $S_{0}^{L}\left[ A_{\mu
}^{a},H_{\mu }^{a},\varphi _{a},B_{a}^{\mu \nu }\right] +g\int d^{2}x\alpha
_{0}+O\left( g^{2}\right) $ stands for a consistent deformation of the free
action, with deformed gauge transformations $\bar{\delta}_{\epsilon }A_{\mu
}^{a}=\partial _{\mu }\epsilon ^{a}+g\beta _{\mu }^{a}+O\left( g^{2}\right)
,\;\bar{\delta}_{\epsilon }H_{\mu }^{a}=\partial ^{\nu }\epsilon _{\mu \nu
}^{a}+\rho _{\mu }^{a}+O\left( g^{2}\right) $, $\bar{\delta}_{\epsilon
}\varphi _{a}=g\beta _{a}+O\left( g^{2}\right) $, $\bar{\delta}_{\epsilon
}B_{a}^{\mu \nu }=\beta _{a}^{\mu \nu }+O\left( g^{2}\right) $, then the
deformed solution to the master equation 
\begin{equation}
\bar{S}=S+g\int d^{2}x\alpha +O\left( g^{2}\right) ,  \label{3}
\end{equation}
satisfies $\left( \bar{S},\bar{S}\right) =0$, where 
\begin{equation}
S=S_{0}^{L}\left[ A_{\mu }^{a},H_{\mu }^{a},\varphi _{a},B_{a}^{\mu \nu
}\right] +\int d^{2}x\left( A_{a}^{*\mu }\partial _{\mu }\eta
^{a}+H_{a}^{*\mu }\partial ^{\nu }\eta _{\mu \nu }^{a}\right) ,  \label{4}
\end{equation}
and $\alpha =\alpha _{0}+A_{a}^{*\mu }\bar{\beta}_{\mu }^{a}+H_{a}^{*\mu }%
\bar{\rho}_{\mu }^{a}+\varphi ^{*a}\bar{\beta}_{a}+B_{\mu \nu }^{*a}\bar{%
\beta}_{a}^{\mu \nu }+`{\rm more}$' ($g$ is the so-called deformation
parameter or coupling constant). The terms $\bar{\beta}_{\mu }^{a}$, $\bar{%
\rho}_{\mu }^{a}$, $\bar{\beta}_{a}$, $\bar{\beta}_{a}^{\mu \nu }$ are
obtained by replacing the gauge parameters $\epsilon ^{a}$ and $\epsilon
_{\mu \nu }^{a}$ respectively with the fermionic ghosts $\eta ^{a}$ and $%
\eta _{\mu \nu }^{a}$ in the functions $\beta _{\mu }^{a}$, $\rho _{\mu }^{a}
$, $\beta _{a}$ and $\beta _{a}^{\mu \nu }$. The fields carrying a star
denote the antifields of the corresponding fields or ghosts. The Grassmann
parity of an antifield is opposite to that of the corresponding field/ghost.
The pure ghost number (${\rm pgh}$) and the antighost number (${\rm antigh}$%
) of the fields, ghosts and antifields are valued like 
\begin{equation}
{\rm pgh}\Phi ^{\alpha _{0}}={\rm pgh}\left( \Phi _{\alpha _{0}}^{*}\right)
=0,\;{\rm pgh}\left( \eta ^{\alpha _{1}}\right) =1,\;{\rm pgh}\left( \eta
_{\alpha _{1}}^{*}\right) =0  \label{5}
\end{equation}
\begin{equation}
{\rm antigh}\Phi ^{\alpha _{0}}=0,\;{\rm antigh}\Phi _{\alpha _{0}}^{*}=1,\;%
{\rm antigh}\left( \eta ^{\alpha _{1}}\right) =0,\;{\rm antigh}\left( \eta
_{\alpha _{1}}^{*}\right) =2,  \label{6}
\end{equation}
where we employed the notations 
\begin{equation}
\Phi ^{\alpha _{0}}=\left( A_{\mu }^{a},H_{\mu }^{a},\varphi _{a},B_{a}^{\mu
\nu }\right) ,\;\Phi _{\alpha _{0}}^{*}=\left( A_{a}^{*\mu },H_{a}^{*\mu
},\varphi ^{*a},B_{\mu \nu }^{*a}\right) ,  \label{7}
\end{equation}
\begin{equation}
\eta ^{\alpha _{1}}=\left( \eta ^{a},\eta _{\mu \nu }^{a}\right) ,\;\eta
_{\alpha _{1}}^{*}=\left( \eta _{a}^{*},\eta _{a}^{*\mu \nu }\right) .
\label{8}
\end{equation}
The BRST\ symmetry of the free theory, $s\bullet =\left( \bullet ,S\right) $%
, simply decomposes as the sum between the Koszul-Tate differential $\delta $
and the exterior derivative along the gauge orbits $\gamma $, $s=\delta
+\gamma $, where the degree of $\delta $ is the antighost number (${\rm %
antigh}\left( \delta \right) =-1$, ${\rm antigh}\left( \gamma \right) =0$),
and that of $\gamma $ is the pure ghost number (${\rm pgh}\left( \gamma
\right) =1$, ${\rm pgh}\left( \delta \right) =0$). The grading of the BRST
differential is named ghost number (${\rm gh}$) and is defined in the usual
manner like the difference between the pure ghost number and the antighost
number, such that ${\rm gh}\left( s\right) =1$. The actions of $\delta $ and 
$\gamma $ on the generators of the BRST complex can be written as 
\begin{equation}
\delta \Phi ^{\alpha _{0}}=0,\;\delta \eta ^{\alpha _{1}}=0,  \label{9}
\end{equation}
\begin{equation}
\delta A_{a}^{*\mu }=\partial _{\nu }B_{a}^{\nu \mu },\;\delta H_{a}^{*\mu
}=-\partial ^{\mu }\varphi _{a},\;\delta \varphi ^{*a}=\partial ^{\mu
}H_{\mu }^{a},  \label{10}
\end{equation}
\begin{equation}
\delta B_{\mu \nu }^{*a}=-\frac{1}{2}\partial _{\left[ \mu \right.
}A_{\left. \nu \right] }^{a},\;\delta \eta _{a}^{*}=-\partial _{\mu
}A_{a}^{*\mu },\;\delta \eta _{a}^{*\mu \nu }=\frac{1}{2}\partial ^{\left[
\mu \right. }H_{a}^{*\left. \nu \right] },  \label{11}
\end{equation}
\begin{equation}
\gamma A_{\mu }^{a}=\partial _{\mu }\eta ^{a},\;\gamma H_{\mu }^{a}=\partial
^{\nu }\eta _{\mu \nu }^{a},\;\gamma \varphi _{a}=\gamma B_{a}^{\mu \nu }=0,
\label{12}
\end{equation}
\begin{equation}
\gamma \eta ^{\alpha _{1}}=\gamma \Phi _{\alpha _{0}}^{*}=\gamma \eta
_{\alpha _{1}}^{*}=0.  \label{13}
\end{equation}

The master equation $\left( \bar{S},\bar{S}\right) =0$ holds to order $g$ if
and only if 
\begin{equation}
s\alpha =\partial _{\mu }j^{\mu },  \label{14}
\end{equation}
for some local $j^{\mu }$. This means that the nontrivial first-order
consistent interactions belong to $H^{0}\left( s|d\right) $, where $d$ is
the exterior space-time derivative. In the case where $\alpha $ is a
coboundary modulo $d$ ($\alpha =s\lambda +\partial _{\mu }b^{\mu }$), then
the deformation is trivial (it can be eliminated by a redefinition of the
fields). In order to investigate the solution to (\ref{14}), we develop $%
\alpha $ according to the antighost number 
\begin{equation}
\alpha =\alpha _{0}+\alpha _{1}+\ldots \alpha _{J},\;{\rm antigh}\left(
\alpha _{k}\right) =k,  \label{15}
\end{equation}
where the last term can be assumed to be annihilated by $\gamma $, $\gamma
\alpha _{J}=0$.\ Thus, we need to know the cohomology of $\gamma $, $H\left(
\gamma \right) $, in order to determine the terms of highest antighost
number in $\alpha $. From (\ref{12}-\ref{13}) it is simple to see that the
cohomology of $\gamma $ is generated by $F_{\mu \nu }^{a}=\partial _{\left[
\mu \right. }A_{\left. \nu \right] }^{a}$, $\partial ^{\mu }H_{\mu }^{a}$, $%
\varphi _{a}$, $B_{a}^{\mu \nu }$, the antifields together with their
derivatives, as well as by the ghosts. If we denote by $e^{M}\left( \eta
^{\alpha _{1}}\right) $ a basis in the space of the polynomials in the
ghosts, it follows that the general solution to the equation $\gamma a=0$
takes the form 
\begin{equation}
a=a_{M}\left( \left[ F_{\mu \nu }^{a}\right] ,\left[ \partial ^{\mu }H_{\mu
}^{a}\right] \left[ \varphi _{a}\right] ,\left[ B_{a}^{\mu \nu }\right]
,\left[ \Phi _{\alpha _{0}}^{*}\right] ,\left[ \eta _{\alpha
_{1}}^{*}\right] \right) e^{M}\left( \eta ^{\alpha _{1}}\right) +\gamma b,
\label{16}
\end{equation}
where the notation $f\left[ q\right] $ signifies that $f$ depends on $q$ and
its derivatives up to a finite order. At this point we recall the cohomology
of $\delta $ modulo the exterior space-time derivative, $H\left( \delta
|d\right) $. On account of the results inferred in \cite{10} it follows that
it is vanishing for all antighost numbers strictly greater than two, $%
H_{J}\left( \delta |d\right) =0$ for $J>2$. Starting from the general form
of an object of antighost number two, $a_{2}=N^{\alpha _{1}}\eta _{\alpha
_{1}}^{*}+M^{\alpha _{0}\beta _{0}}\Phi _{\alpha _{0}}^{*}\Phi _{\beta
_{0}}^{*}$, where $N^{\alpha _{1}}$ and $M^{\alpha _{0}\beta _{0}}$ are
functions of $\Phi ^{\alpha _{0}}$ and their derivatives, and requiring that 
$a_{2}$ belongs to $H_{2}\left( \delta |d\right) $, hence $\delta
a_{2}=\partial _{\mu }m^{\mu }$, we get that, up to a trivial term, the most
general element from $H_{2}\left( \delta |d\right) $ can be represented as 
\begin{eqnarray}
&&a_{2}=K\left( \frac{\delta W}{\delta \varphi _{c}}\eta _{c}^{*}-\frac{%
\delta ^{2}W}{\delta \varphi _{c}\delta \varphi _{d}}B_{d\mu \nu }\eta
_{c}^{*\mu \nu }+\frac{\delta ^{2}W}{\delta \varphi _{c}\delta \varphi _{d}}%
A_{c}^{*\mu }H_{d\mu }^{*}-\right.   \nonumber \\
&&\left. \frac{1}{2}\frac{\delta ^{3}W}{\delta \varphi _{c}\delta \varphi
_{d}\delta \varphi _{e}}B_{d\mu \nu }H_{c}^{*\mu }H_{e}^{*\nu }\right)
+K_{\mu \nu }\left( \frac{\delta U}{\delta \varphi _{c}}\eta _{c}^{*\mu \nu
}+\right.   \nonumber \\
&&\left. \frac{1}{2}\frac{\delta ^{2}U}{\delta \varphi _{c}\delta \varphi
_{d}}H_{c}^{*\mu }H_{d}^{*\nu }\right) =K\tilde{a}_{2}+K_{\mu \nu }\bar{a}%
_{2}^{\mu \nu },  \label{17}
\end{eqnarray}
where $W$ and $U$ are some functions involving only the scalar fields $%
\varphi _{a}$, while $K$ and $K_{\mu \nu }$ represent
some constants, with $K_{\mu \nu }=-K_{\nu \mu }$. From (\ref{17}) we find
in straightforward manner that $\delta \tilde{a}_{2}=\partial _{\mu }\tilde{m%
}^{\mu }$ and $\delta \bar{a}_{2}^{\mu \nu }=\partial _{\beta }\bar{m}%
^{\beta \mu \nu }$. Moreover, $a_{2}$ is $\gamma $-closed, $\gamma a_{2}=0$.
We have enough information as to solve the equation (\ref{14}). Since $%
H_{J}\left( \delta |d\right) $ vanishes for $J>2$, one can assume that $%
\alpha =\alpha _{0}+\alpha _{1}+\alpha _{2}$. As explained in the above, the
general solution to the equation $\gamma \alpha _{2}=0$ is (up to a trivial
term) $\alpha _{2}=\alpha _{M}e^{M}\left( \eta ^{\alpha _{1}}\right) $,
where ${\rm pgh}\left( e^{M}\left( \eta ^{\alpha _{1}}\right) \right) =2$
and ${\rm antigh}\left( \alpha _{M}\right) =2$. Consequently, we have that 
\begin{equation}
\alpha _{2}=\frac{1}{2}\alpha _{ab}\eta ^{a}\eta ^{b}+\alpha _{\;\;ab}^{\mu
\nu }\eta ^{a}\eta _{\mu \nu }^{b}+\frac{1}{2}\alpha _{\;\;ab}^{\mu \nu
\;\lambda \rho }\eta _{\mu \nu }^{a}\eta _{\lambda \rho }^{b},  \label{18}
\end{equation}
where $\alpha _{ab}$, $\alpha _{\;\;ab}^{\mu \nu }$ and $\alpha
_{\;\;ab}^{\mu \nu \;\lambda \rho }$ are $\gamma $-invariant functions of
antighost number two, that should satisfy in addition the symmetry
properties 
\begin{equation}
\alpha _{ab}=-\alpha _{ba},\;\alpha _{\;\;ab}^{\mu \nu }=-\alpha
_{\;\;ab}^{\nu \mu },\;\alpha _{\;\;ab}^{\mu \nu \;\lambda \rho }=-\alpha
_{\;\;ba}^{\lambda \rho \;\mu \nu }.  \label{19}
\end{equation}
Here come in the results connected with $H_{J}\left( \delta |d\right) =0$
for $J>2$. Equation (\ref{14}) projected on antighost number one is locally
expressed by $\delta \alpha _{2}+\gamma \alpha _{1}=\partial _{\mu }n^{\mu }$%
. A necessary condition for the last equation to possess solution (or,
equivalently, for $\alpha _{1}$ to exist) is that the functions $\alpha _{ab}
$, $\alpha _{\;\;ab}^{\mu \nu }$ and $\alpha _{\;\;ab}^{\mu \nu \;\lambda
\rho }$ belong to $H_{2}\left( \delta |d\right) $ 
\begin{equation}
\delta \alpha _{ab}=\partial _{\mu }k_{\;\;ab}^{\mu },\;\delta \alpha
_{\;\;ab}^{\mu \nu }=\partial _{\beta }k_{\;\;ab}^{\beta \mu \nu },\;\delta
\alpha _{\;\;ab}^{\mu \nu \;\lambda \rho }=\partial _{\beta
}k_{\;\;ab}^{\beta \;\mu \nu \;\lambda \rho }.  \label{20}
\end{equation}
The existence of $\alpha _{1}$ demands in addition that the functions $%
k_{\;\;ab}^{\beta \mu \nu }$ and $k_{\;\;ab}^{\beta \;\mu \nu \;\lambda \rho
}$ satisfy the equations 
\begin{equation}
k_{\;\;ab}^{\beta \mu \nu }\partial _{\beta }\eta _{\mu \nu }^{b}=\mu
_{\;\;ab}^{\nu }\partial ^{\mu }\eta _{\mu \nu }^{b},  \label{21}
\end{equation}
\begin{equation}
k_{\;\;ab}^{\beta \;\mu \nu \;\lambda \rho }\partial _{\beta }\left( \eta
_{\mu \nu }^{a}\eta _{\lambda \rho }^{b}\right) =\sigma _{\;\;ab}^{\nu
\;\lambda \rho }\left( \partial ^{\mu }\eta _{\mu \nu }^{a}\right) \eta
_{\lambda \rho }^{b}+\sigma _{\;\;ab}^{\mu \nu \;\rho }\eta _{\mu \nu
}^{a}\left( \partial ^{\lambda }\eta _{\lambda \rho }^{b}\right) ,
\label{22}
\end{equation}
for some $\mu $ and $\sigma $. In other words, only the objects from $%
H_{2}\left( \delta |d\right) $ that fulfill the relations (\ref{21}-\ref{22}%
) are allowed to enter the solution (\ref{18}). On the other hand, the
result (\ref{17}) ensures that we can take 
\begin{eqnarray}
&&\alpha _{ab}=K\left( \frac{\delta W_{ab}}{\delta \varphi _{c}}\eta
_{c}^{*}-\frac{\delta ^{2}W_{ab}}{\delta \varphi _{c}\delta \varphi _{d}}%
B_{d\mu \nu }\eta _{c}^{*\mu \nu }+\frac{\delta ^{2}W_{ab}}{\delta \varphi
_{c}\delta \varphi _{d}}A_{c}^{*\mu }H_{d\mu }^{*}-\right.   \nonumber \\
&&\left. \frac{1}{2}\frac{\delta ^{3}W_{ab}}{\delta \varphi _{c}\delta
\varphi _{d}\delta \varphi _{e}}B_{d\mu \nu }H_{c}^{*\mu }H_{e}^{*\nu
}\right) ,  \label{23}
\end{eqnarray}
\begin{equation}
\alpha _{\;\;ab}^{\mu \nu }=\bar{K}\left( \frac{\delta U_{ab}}{\delta
\varphi _{c}}\eta _{c}^{*\mu \nu }+\frac{1}{2}\frac{\delta ^{2}U_{ab}}{%
\delta \varphi _{c}\delta \varphi _{d}}H_{c}^{*\mu }H_{d}^{*\nu }\right)
,\;\alpha _{\;\;ab}^{\mu \nu \;\lambda \rho }=0,  \label{24}
\end{equation}
where $W_{ab}$ and $U_{ab}$ depend on the same fields respectively 
like $W$ and $U$, $W_{ab}=-W_{ba}$, and $\bar{K}$
is a constant. In this way, we can write down $\alpha _{2}$ like 
\begin{eqnarray}
&&\alpha _{2}=\frac{1}{2}K\left( \frac{\delta W_{ab}}{\delta \varphi _{c}}%
\eta _{c}^{*}-\frac{\delta ^{2}W_{ab}}{\delta \varphi _{c}\delta \varphi _{d}%
}B_{d\mu \nu }\eta _{c}^{*\mu \nu }+\frac{\delta ^{2}W_{ab}}{\delta \varphi
_{c}\delta \varphi _{d}}A_{c}^{*\mu }H_{d\mu }^{*}-\right.   \nonumber \\
&&\left. \frac{1}{2}\frac{\delta ^{3}W_{ab}}{\delta \varphi _{c}\delta
\varphi _{d}\delta \varphi _{e}}B_{d\mu \nu }H_{c}^{*\mu }H_{e}^{*\nu
}\right) \eta ^{a}\eta ^{b}+\bar{K}\left( \frac{\delta U_{ab}}{\delta
\varphi _{c}}\eta _{c}^{*\mu \nu }+\right.   \nonumber \\
&&\left. \frac{1}{2}\frac{\delta ^{2}U_{ab}}{\delta \varphi _{c}\delta
\varphi _{d}}H_{c}^{*\mu }H_{d}^{*\nu }\right) \eta ^{a}\eta _{\mu \nu }^{b}.
\label{25}
\end{eqnarray}
Rigorously speaking, we could have also added the term $\varepsilon _{\mu
\nu }\alpha _{\;\;ab}^{\mu \nu }$ (resulting from the admissible choice $%
K_{\mu \nu }=\varepsilon _{\mu \nu }$, as $\varepsilon _{\mu \nu }$ are the
only covariant antisymmetric constants in two dimensions) to $\alpha _{ab}$.
However, we avoided this term because we intend to construct only those
deformations that are independent of the space-time dimension. The presence
of $\varepsilon _{\mu \nu }\alpha _{\;\;ab}^{\mu \nu }$ would result, at the
level of the deformed action and accompanying gauge structure, in quantities
proportional to $\varepsilon _{\mu \nu }$, and will be therefore omitted. If
we compute $\delta \alpha _{2}$, we consequently deduce that the term of
antighost number one in the first-order deformation of the solution to the
master equation is expressed by 
\begin{eqnarray}
&&\alpha _{1}=K\left( \frac{\delta W_{ab}}{\delta \varphi _{c}}\left(
A_{c}^{*\mu }A_{\mu }^{a}-B^{*a\mu \nu }B_{c\mu \nu }\right) +\frac{\delta
^{2}W_{ab}}{\delta \varphi _{c}\delta \varphi _{d}}H_{d}^{*\nu }B_{c\mu \nu
}A^{a\mu }\right) \eta ^{b}+  \nonumber \\
&&\bar{K}\left( U_{ab}\left( B^{*a\mu \nu }\eta _{\mu \nu }^{b}+\varphi
^{*b}\eta ^{a}\right) -\frac{\delta U_{ab}}{\delta \varphi _{c}}H_{c}^{*\nu
}\left( A^{a\mu }\eta _{\mu \nu }^{b}+H_{\nu }^{b}\eta ^{a}\right) \right) .
\label{26}
\end{eqnarray}
With $\alpha _{1}$ at hand, we determine $\alpha _{0}$ as the solution to
the equation $\delta \alpha _{1}+\gamma \alpha _{0}=\partial _{\mu }l^{\mu }$%
, which actually reads as 
\begin{equation}
\alpha _{0}=\frac{K}{2}\frac{\delta W_{ab}}{\delta \varphi _{c}}B_{c}^{\mu
\nu }A_{\mu }^{a}A_{\nu }^{b}-\bar{K}U_{ab}A^{a\mu }H_{\mu }^{b}.  \label{27}
\end{equation}
Thus, so far we have completely determined the deformation to order $g$, $%
S_{1}=\int d^{2}x\alpha $.

If we denote by $S_{2}=\int d^{2}x\beta $ the second-order deformation, the
master equation $\left( \bar{S},\bar{S}\right) =0$ holds to order $g^{2}$ if
and only if $\Delta =-2s\beta + \partial _{\mu }\theta ^{\mu }$, where $%
\left( S_{1},S_{1}\right) = \int d^{2}x\Delta $. This means that in order to
have a deformation that is consistent to order $g^{2}$, the integrand of $%
\left( S_{1},S_{1}\right) $ should be $s$-exact modulo $d$. This takes place
if and only if 
\begin{equation}
K=\bar{K},\;U_{ab}=W_{ab},  \label{28}
\end{equation}
and also 
\begin{equation}
t_{bcd}\equiv W_{a\left[ b\right. } \frac{\delta W_{\left. cd\right] }}{%
\delta \varphi _{a}}=0.  \label{29}
\end{equation}
Indeed, on account of (\ref{28}) we get that 
\begin{equation}
\Delta =K^{2}\left( t_{bcd}u^{bcd}+ \frac{\delta t_{bcd}}{\delta \varphi _{e}%
}v_{e}^{\;\;bcd}+\frac{\delta ^{2} t_{bcd}}{\delta \varphi _{e}\delta
\varphi _{n}}z_{en}^{\;\;bcd}+\frac{\delta ^{3} t_{bcd}}{\delta \varphi
_{e}\delta \varphi _{n}\delta \varphi _{m}} w_{enm}^{\;\;bcd}\right) ,
\label{30}
\end{equation}
where 
\begin{equation}
u^{bcd}=\left( A^{b\mu }A^{c\nu }- B^{*b\mu \nu }\eta ^{c}\right) \eta _{\mu
\nu }^{d}-\left( A^{b\mu }H_{\mu }^{d}+ \varphi ^{*b}\eta ^{d}\right) \eta
^{c},  \label{31}
\end{equation}
\begin{eqnarray}
&&v_{e}^{\;\;bcd}=\left( B_{e}^{\mu \nu } A_{\mu }^{b}A_{\nu
}^{d}+A_{e}^{*\mu }A_{\mu }^{b} \eta ^{d}\right) \eta ^{c}+\left( B_{e}^{\mu
\nu }B_{\mu \nu }^{*b}-H_{e}^{*\mu } H_{\mu }^{b}-\right.  \nonumber \\
&&\left. \frac{1}{3}\eta _{e}^{*} \eta ^{b}\right) \eta ^{c}\eta ^{d}+\left(
H_{e}^{*\nu }A^{b\mu }-\eta _{e}^{*\mu \nu } \eta ^{b}\right) \eta ^{c}\eta
_{\mu \nu }^{d},  \label{32}
\end{eqnarray}
\begin{eqnarray}
&&z_{en}^{\;\;bcd}=\left( \eta _{e}^{*\mu \nu }B_{n\mu \nu }\eta
^{b}-B_{e\mu \nu }H_{n}^{*\nu }A^{b\mu }- \frac{1}{2}H_{e}^{*\mu
}H_{n}^{*\nu }\eta _{\mu \nu }^{b}-\right.  \nonumber \\
&&\left. A_{e\mu }^{*}H_{n}^{*\mu } \eta ^{b}\right) \eta ^{c}\eta ^{d},
\label{33}
\end{eqnarray}
\begin{equation}
w_{enm}^{\;\;bcd}=\frac{1}{6}B_{e\mu \nu } H_{n}^{*\mu }H_{m}^{*\nu }\eta
^{b}\eta ^{c}\eta ^{d}.  \label{34}
\end{equation}
From (\ref{31}-\ref{34}) one observes that $\Delta $ given in (\ref{30})
cannot be $s$-exact modulo $d$, therefore it should vanish. This happens if
and only if the functions $W_{ab}$ satisfy (\ref{29}), which is nothing but
Jacobi's identity for a nonlinear gauge algebra \cite{13}. In conclusion,
the consistency at order $g^{2}$ implies $S_{2}=0$. Then, the higher-order
deformation equations are identically satisfied if we choose $S_{3}=\cdots
=S_{k}=\cdots =0$.

For concreteness, we work with $K=1$, such that the deformed solution to the
master equation, consistent to all orders in the deformation parameter,
reads as 
\begin{eqnarray}
&&\bar{S}=\int d^{2}x\left( H_{\mu }^{a} D^{\mu }\varphi _{a}+\frac{1}{2}%
B_{a}^{\mu \nu }\bar{F}_{\mu \nu }^{a}+ A_{a}^{*\mu }\left( \partial _{\mu
}\eta ^{a}-g\frac{\delta W_{bc}}{\delta \varphi _{a}}A_{\mu }^{c}\eta
^{b}\right) +\right.  \nonumber \\
&&H_{a}^{*\mu }\left( \partial ^{\nu } \eta _{\mu \nu }^{a}+g\left( \frac{%
\delta W_{bc}}{\delta \varphi _{a}} A^{b\nu }\eta _{\mu \nu }^{c}-\frac{%
\delta W_{bc}}{\delta \varphi _{a}} H_{\mu }^{c}\eta ^{b}+\frac{\delta
^{2}W_{bc}}{\delta \varphi _{a}\delta \varphi _{d}}B_{d\mu \nu }A^{c\nu
}\eta ^{b}\right) \right) -  \nonumber \\
&&g\varphi ^{*a}W_{ab}\eta ^{b}+ gB^{*a\mu \nu }\left( W_{ab}\eta _{\mu \nu
}^{b}-\frac{\delta W_{ab}}{\delta \varphi _{c}}B_{c\mu \nu } \eta
^{b}\right) +  \nonumber \\
&&\frac{g}{2}\left( \frac{\delta W_{ab}}{\delta \varphi _{c}}\eta _{c}^{*}-%
\frac{\delta ^{2}W_{ab}}{\delta \varphi _{c}\delta \varphi _{d}}B_{d\mu \nu
}\eta _{c}^{*\mu \nu }+ \frac{\delta ^{2}W_{ab}}{\delta \varphi _{c}\delta
\varphi _{d}}A_{c}^{*\mu } H_{d\mu }^{*}-\right.  \nonumber \\
&&\left. \frac{1}{2} \frac{\delta ^{3}W_{ab}}{\delta \varphi _{c}\delta
\varphi _{d}\delta \varphi _{e}} B_{d\mu \nu }H_{c}^{*\mu }H_{e}^{*\nu
}\right) \eta ^{a}\eta ^{b}+ g\left( \frac{\delta W_{ab}}{\delta \varphi _{c}%
}\eta _{c}^{*\mu \nu }+\right.  \nonumber \\
&&\left. \left. \frac{1}{2} \frac{\delta ^{2}W_{ab}}{\delta \varphi
_{c}\delta \varphi _{d}}H_{c}^{*\mu } H_{d}^{*\nu }\right) \eta ^{a}\eta
_{\mu \nu }^{b}\right) ,  \label{35}
\end{eqnarray}
where 
\begin{equation}
D^{\mu }\varphi _{a}=\partial ^{\mu } \varphi _{a}+gW_{ab}A^{b\mu }\varphi
_{a},\;\bar{F}_{\mu \nu }^{a}= \partial _{\left[ \mu \right. }A_{\left. \nu
\right] }^{a}+g\frac{\delta W_{bc}}{\delta \varphi _{a}}A_{\mu }^{b}A_{\nu
}^{c}.  \label{36}
\end{equation}

At this stage, we have all the information for identifying the gauge theory
behind our deformation procedure. From the antighost number zero piece in (%
\ref{35}), it follows that the Lagrangian action that describes the deformed
model has the expression 
\begin{equation}
\bar{S}_{0}\left[ A_{\mu }^{a}, H_{\mu }^{a},\varphi _{a},B_{a}^{\mu \nu
}\right] =\int d^{2}x\left( H_{\mu }^{a} D^{\mu }\varphi _{a}+\frac{1}{2}%
B_{a}^{\mu \nu } \bar{F}_{\mu \nu }^{a}\right) ,  \label{37}
\end{equation}
while from the antighost number one components we read the corresponding
deformed gauge transformations 
\begin{equation}
\bar{\delta}_{\epsilon }A_{\mu }^{a}= \partial _{\mu }\epsilon ^{a}-g\frac{%
\delta W_{bc}}{\delta \varphi _{a}} A_{\mu }^{c}\epsilon ^{b},  \label{38}
\end{equation}
\begin{equation}
\bar{\delta}_{\epsilon }H_{\mu }^{a}= \partial ^{\nu }\eta _{\mu \nu
}^{a}+g\left( \frac{\delta W_{bc}}{\delta \varphi _{a}}A^{b\nu }\epsilon
_{\mu \nu }^{c}- \frac{\delta W_{bc}}{\delta \varphi _{a}}H_{\mu
}^{c}\epsilon ^{b}+ \frac{\delta ^{2}W_{bc}}{\delta \varphi _{a}\delta
\varphi _{d}}B_{d\mu \nu } A^{c\nu }\epsilon ^{b}\right) ,  \label{39}
\end{equation}
\begin{equation}
\bar{\delta}_{\epsilon }\varphi _{a}= -gW_{ab}\epsilon ^{b},\;\bar{\delta}%
_{\epsilon }B_{a}^{\mu \nu }= g\left( W_{ab}\epsilon ^{b\mu \nu }-\frac{%
\delta W_{ab}}{\delta \varphi _{c}} B_{c}^{\mu \nu }\epsilon ^{b}\right) .
\label{40}
\end{equation}
The form of the coefficients of the terms proportional with one antifield of
the ghosts and two ghosts indicate that the gauge algebra is nonlinear, and,
moreover, the presence of the quantities involving two antifields of the
original fields and two ghosts shows that this algebra only closes on-shell.
It is now clear that the deformed Lagrangian action, as well as the
resulting gauge structure, does not contain in any way the two-dimensional
antisymmetric symbol, as we have previously required. In view of this, there
is hope that our deformation mechanism can be properly generalized to higher
dimensions.

To conclude with, in this paper we have investigated the consistent
interactions that can be introduced among a set of scalar fields, two types
of one-forms and a system of two-forms in two dimensions, described in the
free limit by an abelian BF theory. Starting with the BRST differential for
the free theory, $s=\delta +\gamma $, we compute the consistent first-order
deformation with the help of some cohomological arguments. Next, we prove
that the deformation is also second-order consistent, and, moreover, matches
the higher-order deformation equations. As a result, we are precisely led to
a two-dimensional nonlinear gauge theory, that can be in principle extended
to higher dimensions. Our deformation procedure modifies the gauge
transformations, as well as their algebra. Moreover, the deformed gauge
algebra is open and closes on-shell.

\section*{Acknowledgment}

This work has been supported by a Romanian National Council for Academic
Scientific Research (CNCSIS) grant.


\begin{thebibliography}{99}
\bibitem{1}  E. S. Fradkin, G. A. Vilkovisky, Phys. Lett. {\bf B55} (1975)
224; I. A. Batalin, G. A. Vilkovisky, Phys. Lett. {\bf B69} (1977) 309; E.
S. Fradkin, T. E. Fradkina, Phys. Lett. {\bf B72} (1978) 343; I. A. Batalin,
G. A. Vilkovisky, Phys. Lett. {\bf B102} (1981) 27; I. A. Batalin, E. S.
Fradkin, Phys. Lett. {\bf B122} (1983) 157; I. A. Batalin, G. A. Vilkovisky,
Phys. Rev. {\bf D28} (1983) 2567; I. A. Batalin, G. A. Vilkovisky, J. Math.
Phys. {\bf 26} (1985) 172; M. Henneaux, Phys. Rep. {\bf 126} (1985) 1; A. D.
Browning, D. Mc Mullan, J. Math. Phys. {\bf 28} (1987) 438; M.
Dubois-Violette, Ann. Inst. Fourier {\bf 37} (1987) 45; D. Mc Mullan, J.
Math. Phys. {\bf 28} (1987) 428; M. Henneaux, C. Teitelboim, Commun. Math.
Phys. {\bf 115} (1988) 213; J. D. Stasheff, Bull. Amer. Math. Soc. {\bf 19}
(1988) 287; J.Fisch, M. Henneaux, J. D. Stasheff, C. Teitelboim, Commun.
Math. Phys. {\bf 120} (1989) 379; M. Henneaux, Nucl. Phys. {\bf B} (Proc.
Suppl) {\bf 18A} (1990) 47; M. Henneaux, C. Teitelboim, Quantization of
Gauge Systems (Princeton University Press, Princeton, New Jersey) 1992

\bibitem{4}  B. Voronov, I. V. Tyutin, Theor. Math. Phys. {\bf 50} (1982)
218; B. Voronov, I. V. Tyutin, Theor. Math. Phys. {\bf 52} (1982) 628; J.
Gomis, S. Weinberg, Nucl. Phys. {\bf B469} (1996) 473; S. Weinberg, The
Quantum Theory of Fields (Cambridge University Press, Cambridge) 1996

\bibitem{5}  O. Piguet, S. P. Sorella, Algebraic Renormalization:
Perturbative Renormalization, Symmetries and Anomalies (Lecture Notes in
Physics, Springer Verlag, Berlin) Vol. {\bf 28} 1995

\bibitem{6}  P. S. Howe, V. Lindstr\H{o}m, P. White, Phys. Lett. {\bf B246}
(1990) 130; W. Troost, P. van Nieuwenhuizen, A. van Proeyen, Nucl. Phys. 
{\bf B333} (1990) 727; G. Barnich, M. Henneaux, Phys. Rev. Lett. {\bf 72}
(1994) 1588; G. Barnich, Mod. Phys. Lett.{\bf A9} (1994) 665; G. Barnich,
Phys. Lett. {\bf B419} (1998) 211

\bibitem{7}  F. Brandt, M. Henneaux, A. Wilch, Phys. Lett. {\bf B387} (1996)
320

\bibitem{7a}  R. Arnowitt, S. Deser, Nucl. Phys. {\bf 49} (1963) 133; J.
Fang, C. Fronsdal, J. Math. Phys. {\bf 20} (1979) 2264; F. A. Berends, G. H.
Burgers, H. Van Dam, Z. Phys. {\bf C24} (1984) 247; F. A. Berends, G. H.
Burgers, H. Van Dam, Nucl. Phys. {\bf B260} (1985) 295; A. K. H. Bengtsson,
Phys. Rev. {\bf D32} (1985) 2031

\bibitem{8}  G. Barnich, M. Henneaux, Phys. Lett. {\bf B311} (1993) 123; J
D. Stasheff, preprint q-alg/9702012; J. D. Stasheff, preprint
hep-th/9712157; J. A. Garcia, B. Knaepen, Phys. Lett. {\bf B441} (1998) 198

\bibitem{11}  C. Bizdadea, E. M. Cioroianu, S. O. Saliu, Class. Quantum
Grav. {\bf 17} (2000) 2007; C. Bizdadea, L. Saliu, S. O. Saliu, Int. J. Mod.
Phys. {\bf A15} (2000) 893; C. Bizdadea, S. O. Saliu, Phys. Scripta {\bf 62}
(2000) 261; C. Bizdadea, preprint hep-th/0003199

\bibitem{9}  E. Cremmer, B. Julia, J. Scherk, Phys. Lett. {\bf B76} (1978)
409; R. Wald, Phys. Rev. {\bf D33} (1986) 3613; G. Barnich, M. Henneaux, R.
Tatar, Int. J. Mod. Phys. {\bf D3} (1994) 139; G. Barnich, F. Brandt, M.
Henneaux, Commun. Math. Phys. {\bf 174} (1995) 93; G. Barnich, F. Brandt, M.
Henneaux, Nucl. Phys. {\bf B455} (1995) 357; M. Henneaux, Phys. Lett. {\bf %
B368} (1996) 83; M. Henneaux, B. Knaepen, Phys. Rev. {\bf D56} (1997) 6076;
M. Henneaux, B. Knaepen, C. Schomblond, Lett. Math. Phys. {\bf 42} (1997)
337; M. Henneaux, V. E. R. Lemes, C. A. G. Sasaki, S. P. Sorella, O. S.
Ventura, I. C. Q. Vilar, Phys. Lett. {\bf B410} (1997) 195; F. Brandt, Ann.
Phys. (N.Y.) {\bf 259} (1997) 253; X. Bekaert, M. Henneaux, Int. J. Theor.
Phys. {\bf 38} (1999) 1161; X. Bekaert, M. Henneaux, A. Sevrin, Phys. Lett. 
{\bf B468} (1999) 228; C. Bizdadea, M. G. Mocioac\u {a}, S. O. Saliu, Phys.
Lett. {\bf B459} (1999) 145; X. Bekaert, M. Henneaux, A. Sevrin, Nucl. Phys. 
{\bf B} (Proc. Suppl.) {\bf 88} (2000) 27; K. I. Izawa, Prog. Theor. Phys. 
{\bf 103} (2000) 225; C. Bizdadea, L. Saliu, S. O. Saliu, Phys. Scripta {\bf %
61} (2000) 307; X. Bekaert, M. Henneaux, A. Sevrin, preprint hep-th/0004049;
N. Boulanger, T. Damour, L. Gualtieri, M. Henneaux, preprint hep-th/0007220;
N. Ikeda, preprint hep-th/0010096

\bibitem{10}  M. Henneaux, B. Knaepen, C. Schomblond, Commun. Math. Phys. 
{\bf 186} (1997) 137

\bibitem{12}  for a review, D. Birmingham, M. Blau, M. Rakowski, G.
Thompson, Phys. Rep. {\bf 209} (1991) 129

\bibitem{13}  N. Ikeda, K. I. Izawa, Prog. Theor. Phys. {\bf 89} (1993) 223;
N. Ikeda, K. I. Izawa, Prog. Theor. Phys. {\bf 89} (1993) 1077; N. Ikeda, K.
I. Izawa, Prog. Theor. Phys. {\bf 90} (1993) 237; N. Ikeda, Ann. Phys. {\bf %
235} (1994) 235; P. Schaller, T. Strobl, Phys. Lett. {\bf B337} (1994) 266;
P. Schaller, T. Strobl, Proceedings ``Dubna 1994--- Finite Dimensional
Integrable Systems'' (1994) 181; P. Schaller, T. Strobl, in Lecture Notes in
Physics, Vol. {\bf 469} (Springer Verlag, Berlin) (1996) 321; T. Kl\"{o}sch,
P. Schaller, T. Strobl, Helv. Phys. Acta {\bf 69} (1996) 305; W. Kummer, H.
Liebl, Vassilevich, Nucl. Phys. {\bf B493} (1997) 491

\bibitem{14}  D. J. Gross, A. A. Migdal, Nucl. Phys. {\bf B340} (1990) 333;
T. Banks, M. O'Loughlin, Nucl. Phys. {\bf B362} (1991) 649; E. Witten,
Proceedings ``New York 1991--- Differential Geometry Methods in Theoretical
Physics'', Vol {\bf 1} (1991) 176; G. Grignani, G. Nardelli, Class. Quantum
Grav. {\bf 10} (1993) 2569; T. Fujiwara, Y. Igaraschi, J. Kubo, T. Tabei,
Phys. Rev. {\bf D48} (1993) 1736; T. Fujiwara, T. Tabei, Y. Igaraschi, K.
Maeda, J. Kubo, Mod. Phys. Lett. {\bf A8} (1993) 2147; T. Strobl, Int. J.
Mod. Phys. {\bf D3} (1994) 281; R. Jackiw, preprint gr-qc/9511048

\bibitem{15}  K. Schoutens, A. Sevrin, P. van Nieuwenhuizen, Commun. Math.
Phys. {\bf 124} (1989) 87; S. Hirano, Y. Kazama, Y. Satoh, Phys. Rev. {\bf %
D48} (1993) 1687; \"{O}. Dayi, Mod. Phys. Lett. {\bf A9} (1994) 2157; D.
Louis-Martinez, G. Kunstatter, Phys. Rev. {\bf D52} (1995) 3494; M.
Blagojevi\'{c}, M. Vasili\'{c}, T. Vuka\v {s}inac, Class. Quantum Grav. {\bf %
13} (1996) 3003; \"{O}. Dayi, Int. J. Mod. Phys. {\bf A12} (1997) 4387; P.
M. Lavrov, P. Y. U. Moshin, Class. Quantum Grav. {\bf 16} (1999) 2247; W.
Kummer, G. Tieber, Phys. Rev. {\bf D59} (1999) 044001; D. Grumiller, D.
Hofmann, W. Kummer, preprint gr-qc/0005098
\end{thebibliography}
\end{document}